# Progress and tests on the Instrument Control Electronics for SOXS

M. Colapietro*[a], G. Capasso[a], S. D'Orsi[a], P. Schipani[a], L. Marty[a], S. Savarese[a], I. Coretti[b],
S. Campana[c], R. Claudi[d], M. Aliverti[c], A. Baruffolo[d], S. Ben-Ami[e,f], F. Biondi[g], R. Cosentino[h,i],
F. D'Alessio[j], P. D'Avanzo[c], O. Hershko[e], H. Kuncarayakti[k,l], M. Landoni[c], M. Munari[i],
G. Pignata[m,n], A. Rubin[o], S. Scuderi[p,i], F. Vitali[j], D. Young[q], J. Achrén[r], J.A. Araiza-Duran[s,n],
I. Arcavi[t], A. Brucalassi[m,u], R. Bruch[e], E. Cappellaro[d], M. Della Valle[a], M. De Pascale[d],
R. Di Benedetto[i], A. Gal-Yam[e], M. Genoni[c], M. Hernandez[h], J. Kotilainen[l,k], G. Li Causi[v],
S. Mattila[k], K. Radhakrishnan[d], M. Rappaport[e], D. Ricci[d], M. Riva[c], B. Salasnich[d], S. Smartt[q],
R. Zanmar Sanchez[i], M. Stritzinger[w], H. Ventura[h]

[a]INAF - Osservatorio Astronomico di Capodimonte, Salita Moiariello 16, I-80131 Napoli, Italy
[b]INAF - Osservatorio Astronomico di Trieste, Via G.B. Tiepolo 11, I-34143 Trieste, Italy
[c]INAF - Osservatorio Astronomico di Brera, Via Bianchi 46, I-23807 Merate (LC), Italy
[d]INAF - Osservatorio Astronomico di Padova, Vicolo dell'Osservatorio 5, I-35122 Padova, Italy
[e]Weizmann Institute of Science, Herzl St 234, Rehovot, 7610001, Israel
[f]Harvard-Smithsonian Center for Atrophysics, Cambridge, USA
[g]Max-Planck-Institut für Extraterrestrische Physik, Giessenbachstr. 1,D-85748 Garching, Germany
[h]FGG-INAF, TNG, Rambla J.A. Fernández Pérez 7, E-38712 Breña Baja (TF), Spain
[i]INAF - Osservatorio Astrofisico di Catania, Via S. Sofia 78 30, I-95123 Catania, Italy
[j]INAF - Osservatorio Astronomico di Roma, Via Frascati 33, I-00078 Monte Porzio Catone, Italy
[k]Tuorla Observatory, Department of Physics and Astronomy, FI-20014 University of Turku, Finland
[l]Finnish Centre for Astronomy with ESO (FINCA), FI-20014 University of Turku, Finland
[m]Universidad Andres Bello, Avda. Republica 252, Santiago, Chile
[n]Millennium Institute of Astrophysics (MAS), Santiago, Chile
[o]ESO, Karl Schwarzschild Strasse 2, D-85748, Garching bei München, Germany
[p]INAF - Istituto di Astrofisica Spaziale e Fisica Cosmica, Via Corti 12, I-20133 Milano, Italy
[q]Queen's University Belfast, County Antrim, BT7 1NN, Belfast, UK
[r]Incident Angle Oy, Capsiankatu 4 A 29, FI-20320 Turku, Finland
[s]Centro de Investigaciones en Optica A.C., 37150 León, Mexico
[t]Tel Aviv University, Tel Aviv, Israel
[u]INAF - Osservatorio Astrofisico di Arcetri, Largo E. Fermi 5, I-50125 Firenze, Italy
[v]Istituto di Astrofisica e Planetologia Spaziali, Via Fosso del Cavaliere, I-00133 Roma, Italy
[w]Aarhus University, Ny Munkegade 120, D-8000, Aarhus, Denmark

## ABSTRACT

The forthcoming SOXS (Son Of X-Shooter) will be a new spectroscopic facility for the ESO New Technology Telescope in La Silla, focused on transient events and able to cover both the UV-VIS and NIR bands. The instrument passed the Final Design Review in 2018 and is currently in manufacturing and integration phase. This paper is focused on the assembly and testing of the instrument control electronics, which will manage all the motorized functions, alarms, sensors, and electric interlocks. The electronics is hosted in two main control cabinets, divided in several subracks that are assembled to ensure easy accessibility and transportability, to simplify test, integration and maintenance. Both racks are equipped with independent power supply distribution and have their own integrated cooling systems. This paper shows the assembly strategy, reports on the development status and describes the tests performed to verify the system before the integration into the whole instrument.

*mirko.colapietro@inaf.it; phone +39 081 5575458

**Keywords:** SOXS; New Technology Telescope; control electronics; PLC; motorized stage; control cabinet; subrack

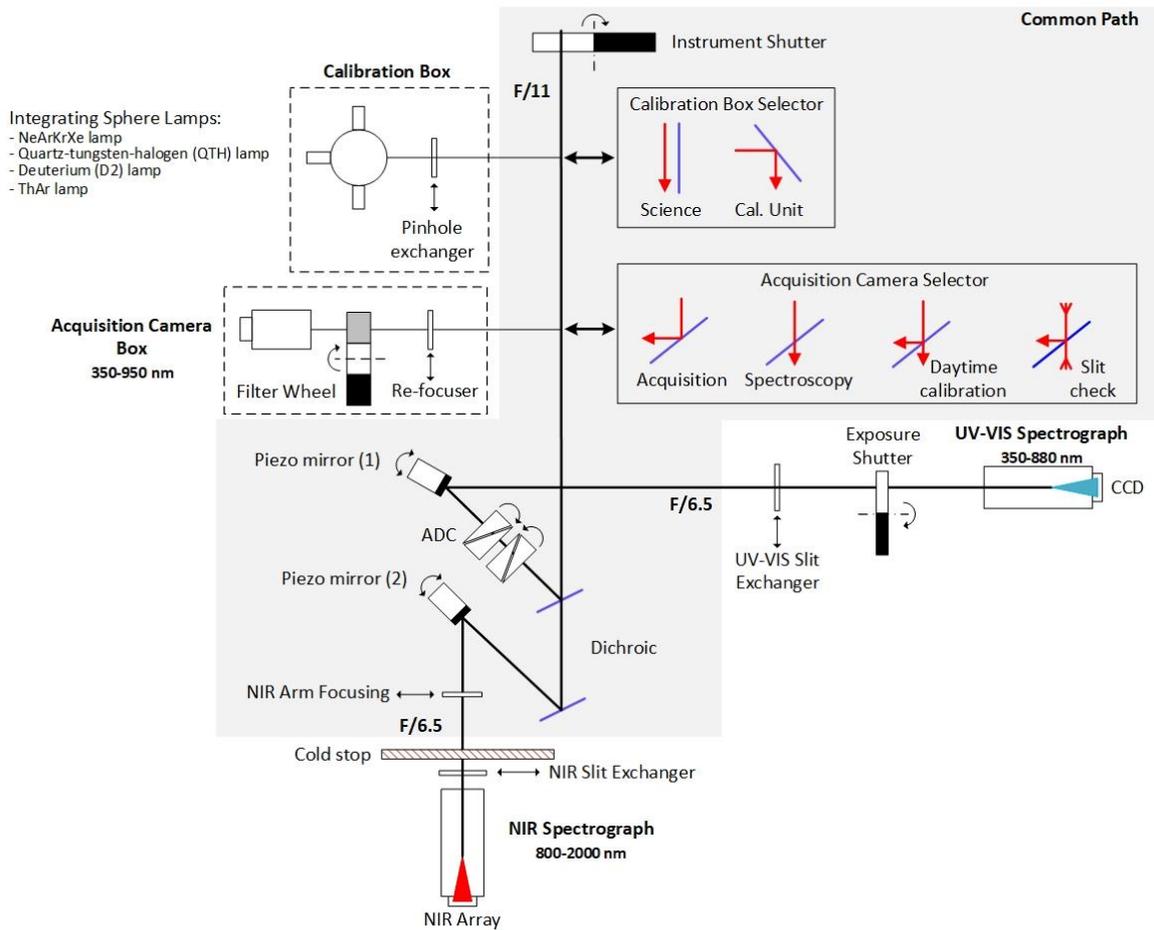

Figure 1. SOXS instrument concept: the Common Path (in grey) splits the light coming from the telescope into the UV-VIS and NIR arms that end with the two spectrographs; the Calibration Box and the Acquisition Camera, connected to the Common Path, can be used for different purposes through the corresponding selector stages.

## 1. INTRODUCTION

SOXS (Son Of X-Shooter) is a new spectrograph designed for the ESO NTT in the La Silla observatory, with the aim of studying transient events simultaneously in the UV-VIS and NIR bands. The project passed the Final Design Review in 2018 and is now in the manufacturing and assembly phase[1].

The SOXS[2] main subsystems are shown in a functional diagram in Figure 1: in normal spectroscopic observation mode, the light enters through the instrument shutter into the Common Path[3],[4], i.e. the central structure of the instrument, where it is splitted into the two UV-VIS[5],[6] and NIR[7],[8] spectrographs. A Calibration Box[9] and an Acquisition Camera[10] are used for wavelength calibration, acquisition, light imaging and engineering purposes (e.g. monitoring of the spectrographs co-alignment).

All the motorized functions, sensors, interlocks and calibration lamps of the instrument are controlled by the Instrument Control Electronics[11] (ICE), which is based on a Programmable Logic Controller (PLC) architecture and distributed in independent subracks to simplify assembly, test and integration. Additional control systems are needed to operate the Cryo-Vacuum subsystem and the UV-VIS[12],[13] and NIR detectors.

After an overview of the overall architecture and the cabinet organization, this paper details the assembly strategy and the tests carried out to verify the correct operation of the control electronics.

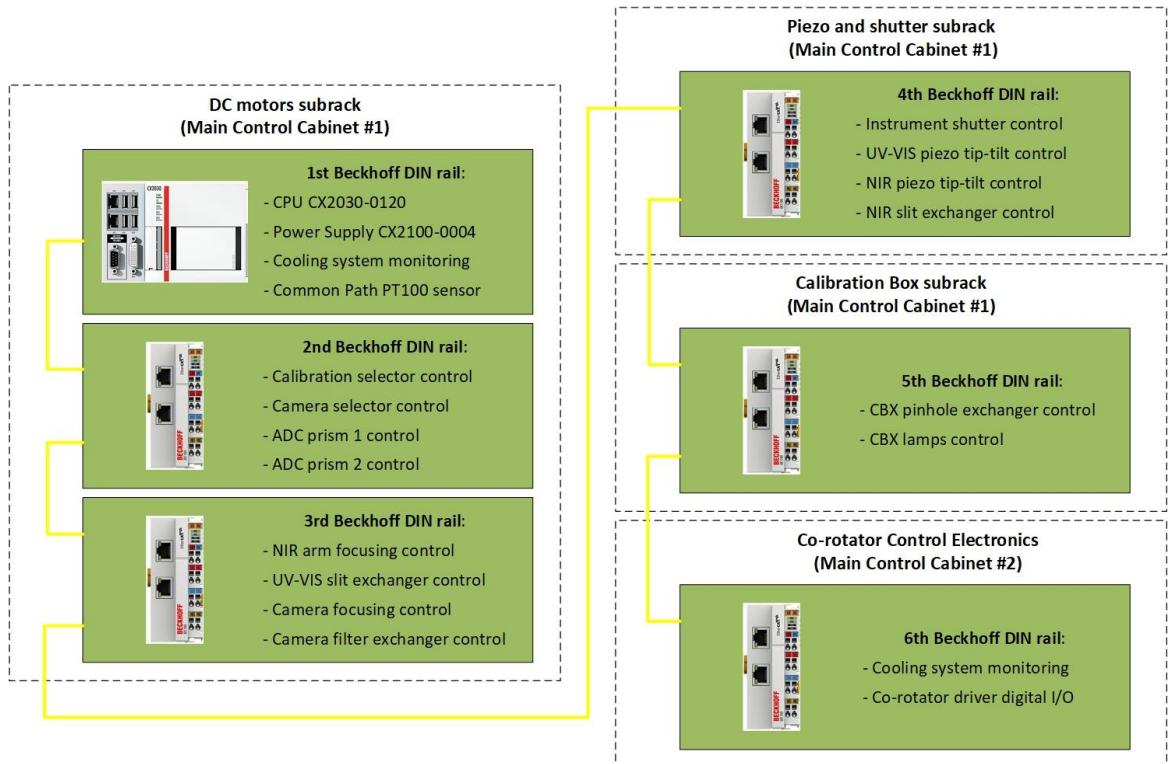

Figure 2. SOXS functions distribution between subracks: a first sub-unit hosts the main CPU and the PLC modules to control the instrument functions based on DC motors; a second one contains all the dedicated controllers for the piezo stages and the instrument shutter; the last two sub-units are dedicated to the electronics of the Calibration Box and the co-rotator control system. All the sub-units are connected together through an EtherCAT fieldbus network.

## 2. SOXS ELECTRONICS OVERVIEW

The Instrument Control Electronics is based on a Beckhoff PLC to control the instrument functions. The PLC has a number of I/O modules to interface with motor drivers, encoders and all other devices; also, it extends the I/O capabilities through an EtherCAT fieldbus network. Moreover, an Open Platform Communication - Unified Architecture (OPC-UA) server is installed on the PLC: it stores all the process variables accessible by the Instrument Software[14] (INS), based on the VLT Common Software standards and running on a Linux Instrument Workstation (IWS), that sends commands and reads instrument telemetries to/from the PLC through the OPC-UA protocol[15].

Such architecture follows the ESO design guidelines, in order to reduce costs, development time and maintenance, and to improve reliability and compatibility with the observatory standards. The validity of the architectural solutions adopted for SOXS has been proved by recent instruments developed for ESO telescopes in the last few years (e.g. ESPRESSO). Electromagnetic compatibility, safety issues, and accessibility for maintenance purposes have been the other fundamental guidelines.

The system relies on Commercial Off-The-Shelf (COTS) industrial components wherever feasible. Custom components have been kept to a minimum, e.g. signal conditioning boards, cables and other minor components. A brief description of the main electronic components is sketched in the following.

*PLC*

The Beckhoff CX2030 CPU is based on the Microsoft Windows Embedded Standard 7 P 32 bit with TwinCAT 3 runtime (XAR). The TwinCAT 3 NC suite enables the real-time control, the control loop tuning and the point to point positioning software, while an OPC-UA server allows for the remote control. I/O modules and drivers are connected to the CPU by means of the EtherCAT fieldbus, enabling an easy decentralization of the functions and the development of a modular architecture.

Figure 2 shows a detailed overview of the function distribution between the subracks, as well as the interconnections between the CPU and the decentralized modules.

*Movement functions*

Most of motorized stages are DC motors, either linear or rotary. Each stage is equipped with an incremental encoder and switches, for limits and homing.

An ESO software library based on TwinCAT 3 provides position and motion control capabilities for such kind of stages. Sensor data are acquired through digital input for switches and differential RS422 interface for encoders, whereas a control module generates the voltage to be applied to the motor[11].

*Cryogenic movement function*

A linear cryogenic piezo stage is located inside the NIR vacuum vessel. A dedicated Micronix controller controls the stage positioning. Commands are sent from the PLC through a RS-485 serial connection[16].

*Tip-Tilt flexure compensators*

A flexure compensation system is present in both UV-VIS and NIR arms to compensate for mechanical flexures. Two dedicated controllers drive two pairs of tip-tilt Physik Instrumente piezo actuators. The main PLC manages the tip-tilt positioning, sending commands through a RS-232 serial connection[16].

*Calibration box*

The calibration unit electronics, i.e. power supplies, security check, failure detection and motion control, are assembled in a dedicated 19" subrack[9]. The connection to the PLC is done through the EtherCAT fieldbus. The modular architecture allowed for the assembly and testing of the calibration subsystem independently of the rest of electronics.

*Co-rotator movement*

A double wheel, driven by a servo motor, drives all cables and pipes connected to the instrument[17],[18]. A dedicated control system[11] makes sure all the cables follow the instrument movement for compensating the rotation of the field. A PLC digital I/O module enables and monitors the status of the motor drive; a signal conditioning board allows to get analogue signals from linear sensors, pushed during the instrument rotation; a safety switch disables the drive in case of control system failures.

*Temperature controllers*

The cabinets are liquid cooled, provided with ESO cooling controllers that manage the coolant flow to minimize the temperature misalignment with respect to the room temperature.

*Cryo-vacuum control system*

A Siemens S7-1500 PLC is hosted in a 19'' subrack to control all the cryo-vacuum electronics and the temperature inside the spectrograph. An independent Lakeshore 336 temperature controller manages the temperatures of the detectors with PID control loops.

*Detector systems*

The UV-VIS and NIR detectors are controlled through two standard ESO NGC controllers, placed onboard the instrument: their power supply units and Linux-based Local Control Units (LLCUs) are hosted in the main control cabinets installed on the Nasmyth platform.

*Cabling*

The cables are designed to optimize the routing, taking into account the following constraints:
- the distance between electronic controllers and the instrument devices is in the 8-10 m range;
- the Beckhoff TTL input modules are sensitive at high frequencies (input filter set at 0.05μsec), so cables are designed to reduce high frequency noise.

All cables and connectors are labeled according to a well-defined policy, to simplify the maintenance.

*Racks and cabinets*

The two 19'' racks on the instrument platform are seismic resistant and liquid cooled. They host almost all the electronics but the two NGCs and some power supply units (PSUs), placed on the instrument because of cable length limitations. Figure 3 shows their design.

The first cabinet (on the left) hosts, from top to bottom: the two ESO NGC power supplies, the subrack with the piezo stages and the instrument shutter controllers, the main PLC subrack with the Beckhoff CPU and the local control panel, the Calibration Box subrack for the calibration lamps.

The second cabinet (on the right) hosts, from top to bottom: the two ESO NGC LLCUs, the co-rotator control system, the Cryo-vacuum control system subrack, the dedicated temperature controller for the detectors.

The modular division in subracks allows for the splitting of the control electronics in different pre-integration sites, as needed in the early phases of the project.

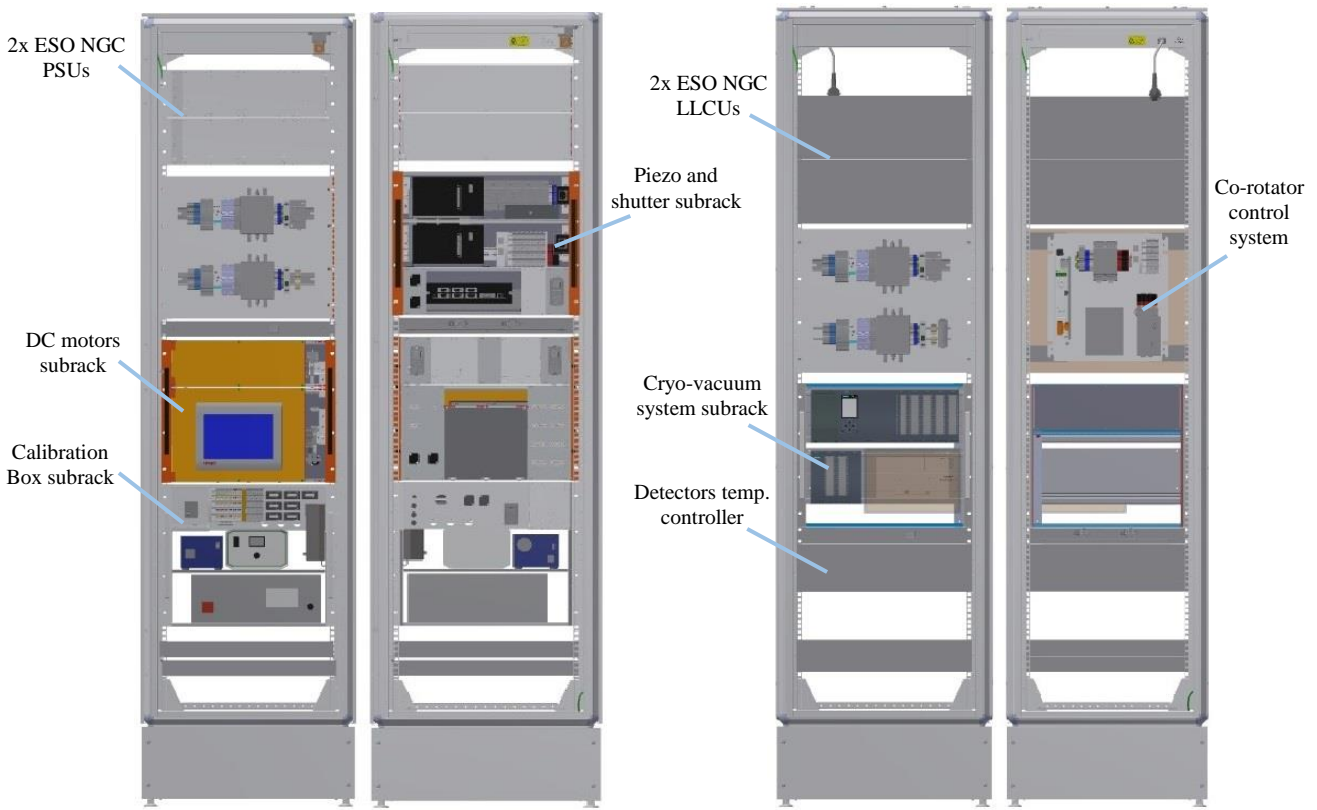

Figure 3. Front and rear view of the first (on the left) and the second (on the right) main control cabinets: both the cabinets are equipped with a panel for the internal power distribution, an Ethernet switch, an additional 1HE fans unit and a 1HE ESO Thermal Control Unit for the internal temperature management.

## 3. ELECTRONICS ASSEMBLY

### 3.1 Main control cabinets

The two SOXS cabinets are nVent Schroff Varistar LHX3 type (2200 mm high, 600 mm large and 800 mm deep). They are equipped with steel doors, lifting eyes on the top cover, adjustable feet, a heat exchanger in the base plinth and a fan tray in the top cover: the cold air flows evenly through the cabinet, while the reflowing of the warm air to the heat exchanger occurs through the channels in the side panels.

Both cabinets host an Ethernet switch and get UPS and normal single-phase power supply from the observatory. A main switch is installed on the front doors for each section and an inner panel provides the internal power distribution (Figure 4).

The plates are fixed on the top of the cabinets and are divided in two sections, for the UPS and normal power, mounted with standard DIN rails equipped with:

- a surge protection device;
- a residual current circuit breaker with overcurrent protection;

- an over-temperature sensing circuit based on a thermostat switch and an under-voltage release that cuts the power supply if a temperature threshold is exceeded;
- a mains electromagnetic compatible (EMC) RFI filter;
- a switch to power off the elements powered by the internal Schuko multisockets.

A 1HE ESO Thermal Control Unit is installed in each cabinet to manage the internal temperature, door switches and the coolant liquid flow. Figure 5 shows the cooling control system architecture under finalization: the Thermal Control Unit monitors cabinet, ambient and coolant temperature and controls a hydraulic valve to adjust the coolant flow into the heat exchanger, setting the cabinet internal temperature within the desired range.

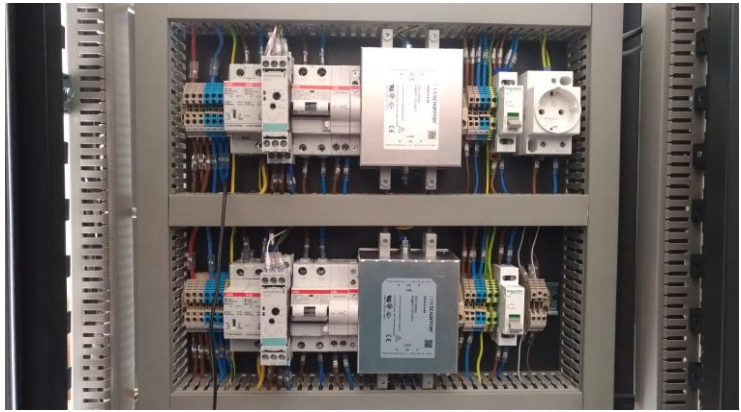

Figure 4. Electrical components for the power distribution circuits into the cabinets. Two sections can be identified: above, the one for the normal power distribution; below, the one for the UPS power distribution.

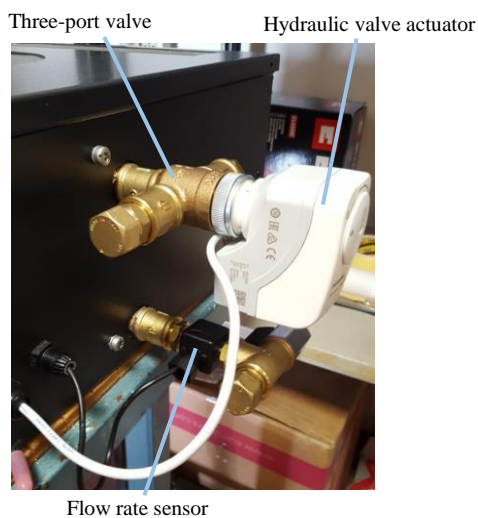

Figure 5. Cooling control system for managing the temperature into the cabinets.

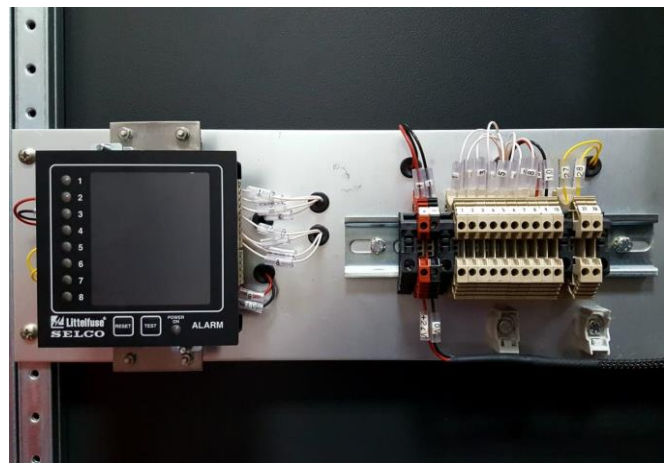

Figure 6. Alarm monitoring system: the alarms are collected by the programmable digital alarm panel (left) that alerts the central alarm system in case of issues.

A SELCO alarm monitoring system (Figure 6) collects all the alarm signals from Cryo-Vacuum subsystem, alerting the observatory alarm system in the case of issues requiring a human intervention.

## 3.2 DC motors subrack

Figure 7 shows the front and the rear views of the 9U DC motor control subrack during assembly and test. On the front side (left), a control panel enables a local control for maintenance purpose only, using engineering interfaces based on TwinCAT software and ESO libraries. On the back side (right), the power supply units are visible, as well as all the external connectors.

The subrack hosts three DIN rails: one supports the Beckhoff CPU and the modules for monitoring temperatures and the ESO Thermal Control Unit; the other two rails (Figure 8) include the modules to control 8 out of 9 SOXS motorized functions based on DC motor + encoder (the last one, the Calibration pinhole exchanger, is controlled by modules installed in the Calibration Box subrack).

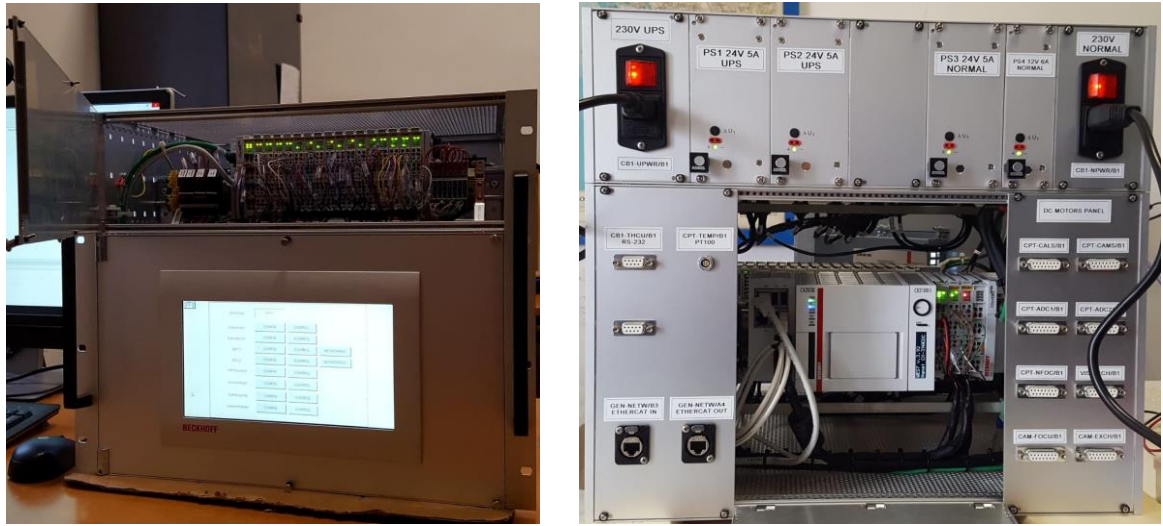

Figure 7. Front and rear view of the DC motors subrack: on the front side, the control panel allows the local control and monitoring of the instrument functions; on the back side, the Beckhoff CPU, the power supplies and all the external interfaces are accessible.

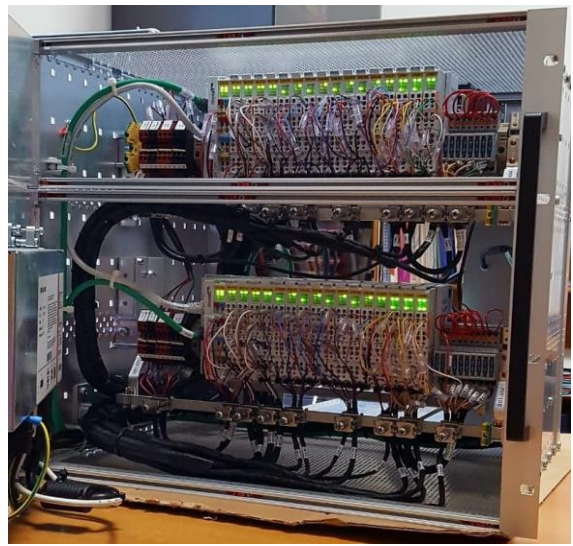

Figure 8. Beckhoff DIN rails for the DC motors control: the upper one controls the $24V_{DC}$ motors; the lower one controls the $12V_{DC}$ motors. Both the rails start with an EtherCAT coupler that ensure the communication with the CPU. Pull-up resistors (at the end of the rails) on limit switch signals prevent issues in case of wire breakage.

### 3.3 Piezo and shutter subrack

Figure 9 shows the structure of the 9U piezo and shutter control subrack: it hosts the four dedicated controllers to control the piezo actuators (the two tip-tilt and the cryogenic linear stage for the NIR spectrograph) and the instrument shutter, as well as all related power supply units.

The controllers interface with the PLC through digital I/O or serial communication. A dedicated DIN rail hosts the EtherCAT coupler and the modules required for this unit.

Two panels, one for the EtherCAT cables and one for the power supply, provide the connection with the external devices. The signal cables coming from the piezo actuators are connected directly to the controllers.

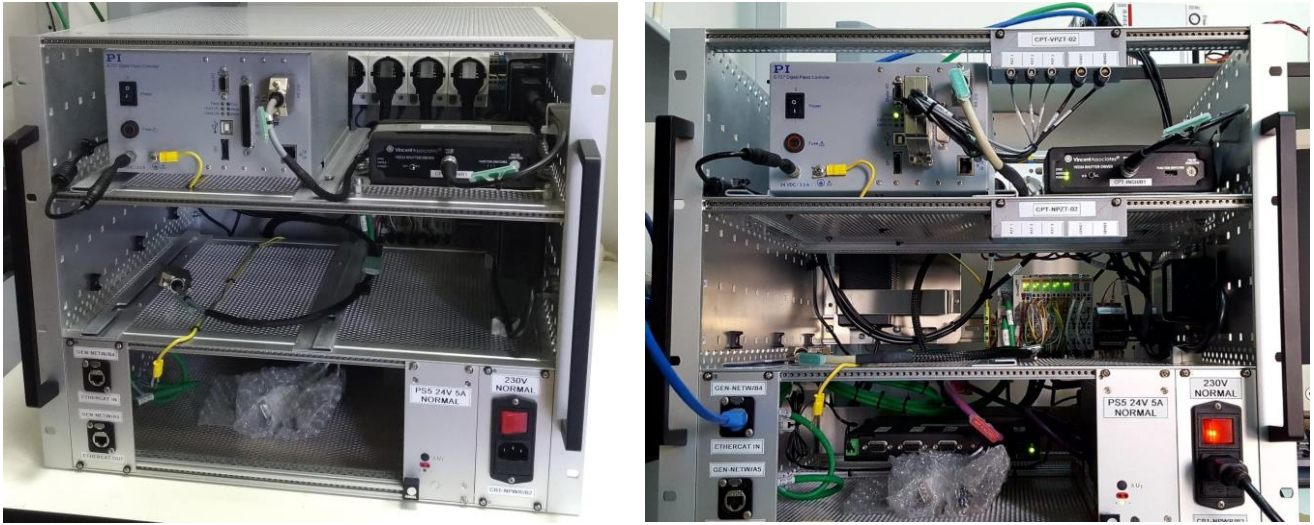

Figure 9. Front view of the Piezo and shutter subrack: on the top, the piezo tip-tilt (left) and the instrument shutter (right) controllers; on the bottom, the cryogenic piezo linear stage controller.

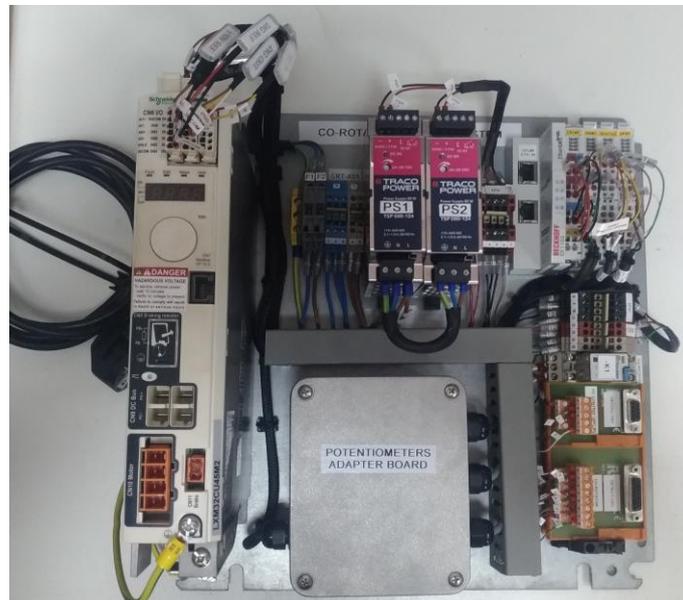

Figure 10. The co-rotator control electronics: all the components are installed on a steel mounting plate that is hosted inside the cabinet. The drive (left) sets the speed and direction of rotation of the motor based on the analog control signal coming from the adapter board.

### 3.4 Co-rotator control system

The co-rotator control system design is described in [2]. The main system components are visible in Figure 10:
- the motor drive for the cable chain;
- the adapter board to manage the feedback coming from the two potentiometers, producing the control signal to the motor;
- the Beckhoff digital I/O modules to enable/disable, reset and monitor the drive;
- the power supplies.

A DIN relay is powered by the "Active" signal coming from the drive: this signal is high (+24V) when the drive is enabled and ready to control the motor, while it is low (0V) if the drive is not enabled or if it is in fault condition. The relay output contact will be inserted into the NTT interlock chain in La Silla, disabling the instrument rotator when the co-rotator is not operational.

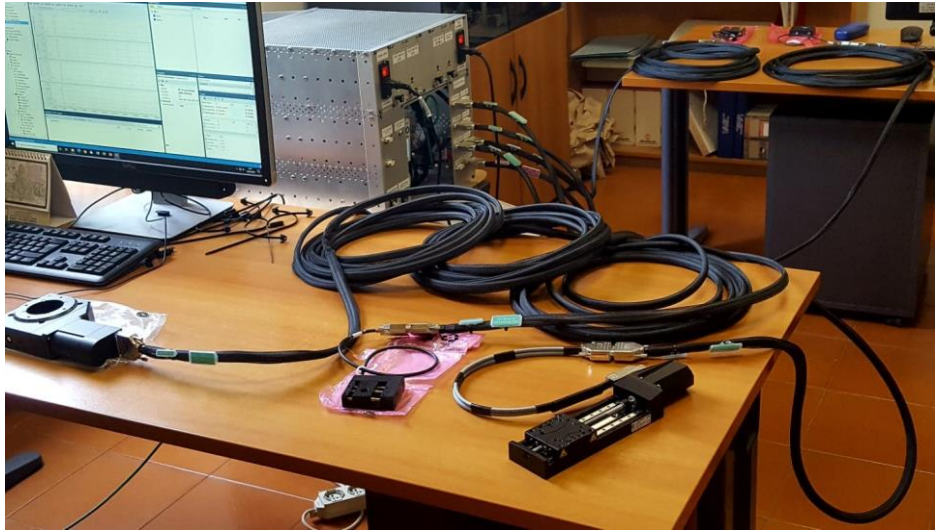

Figure 11. Test moving several stages simultaneously via OPC-UA communication protocol. All the SOXS motorized functions have been successfully tested, together with their own extension cables.

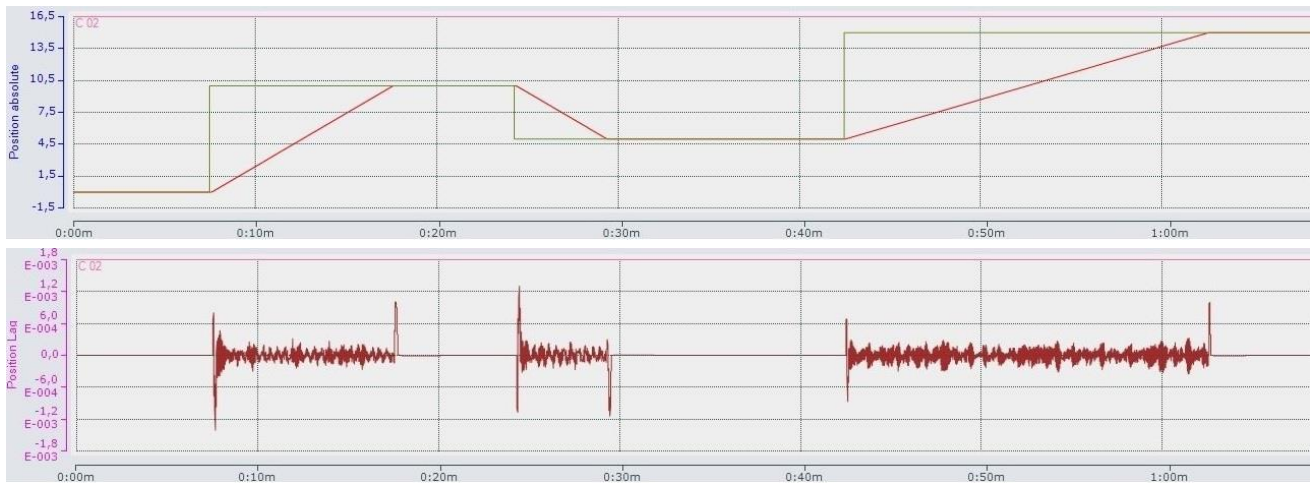

Figure 12. Details on stage positioning: target vs. actual position (top); position lag (bottom). Three consecutive positioning commands are shown: the motor works properly and always reaches the set target position (in green).

## 4. TESTS

Some preliminary tests have been carried out to verify the correct operation, before the integration with the instrument subsystems.
The positioning of all motors has been tested individually, first by using the PLC GUIs supplied with the ESO libraries, then sending commands through the OPC-UA communication protocol, using the UA Expert client from a PC. Parallel tests (Figure 11), moving several stages simultaneously, have been performed as well, revealing no interference between the signals and enabling a preliminary measure of the power consumption.

All the motor parameters (axis, encoder and drive parameters) have been configured within the TwinCAT environment and a first tuning of the PID parameters has been done for each stage, using the Ziegler-Nichols method. A new tuning operation is foreseen when the control system will be connected to the real hardware. Several initialization sequences and positioning commands have been tested to check all parameters, exploiting the Beckhoff TwinCAT Scope View function that allows the real-time monitoring of the PLC variables.

Figure 12 shows an example of the positioning diagram for a linear stage. In the first diagram, the target position, set from the PLC local control interface, is in green while the actual position, read by the encoder, is in red. The second diagram shows that the position lag is always very close to zero (at most between ±0.0012 mm) during the motion and goes to zero when the target position is reached.

## 5. CONCLUSIONS

The SOXS Instrument Control Electronics is based on a Beckhoff PLC system that manages all the motorized functions, sensors, interlocks and calibration lamps. The assembly and tests of the two main control cabinets and subsystems have been carried out at the INAF-Astronomical Observatory of Capodimonte (Naples). All tests have been passed successfully with no relevant issues reported.

The next step will be the forthcoming integration with the instrument opto-mechanics and the Instrument Software. Due to COVID-19 emergency and all the related travel restrictions, we are adopting strategies for risk mitigation: accurate check of parts before shipping to the integration sites, implementation of remote environments for control tuning.

## REFERENCES

[1] P. Schipani, et al., "Development status of the SOXS spectrograph for the ESO-NTT telescope", *Proc. SPIE* 11447 (2020).
[2] P. Schipani, et al., "SOXS: a wide band spectrograph to follow up transients", *Proc. SPIE* 10702, 107020F (2018).
[3] R. Claudi, et al., "The Common Path of SOXS (Son of X-Shooter)", *Proc. SPIE* 10702, 107023T (2018).
[4] F. Biondi, et al., "The AIV strategy of the Common Path of Son of X-Shooter", *Proc. SPIE* 11447 (2020).
[5] A. Rubin, et al., "MITS: the Multi-Imaging Transient Spectrograph for SOXS", *Proc. SPIE 10702*, 107022Z (2018).
[6] A. Rubin, et al., "Progress on the UV-VIS arm of SOXS", *Proc. SPIE* 11447 (2020).
[7] F. Vitali, et al., "The NIR Spectrograph for the new SOXS instrument at the ESO-NTT", *Proc. SPIE* 10702, 1070228 (2018).
[8] F. Vitali, et al., "The development status of the NIR spectrograph for the new SOXS instrument at the NTT", *Proc. SPIE* 11447 (2020).
[9] H. Kuncarayakti, et al., "Design and development of the SOXS calibration unit", *Proc. SPIE* 11447 (2020).
[10] A. Brucalassi, et al., "Final Design and development status of the Acquisition and Guiding System for SOXS", *Proc. SPIE* 11447 (2020).
[11] G. Capasso, et al., "SOXS Control Electronics Design", *Proc. SPIE* 10707, 107072H (2018).
[12] R. Cosentino, et al., "The VIS detector system of SOXS", *Proc. SPIE* 10702, 107022J (2018).
[13] R. Cosentino, et al., "Development status of the UV-VIS detector system of SOXS for the ESO-NTT telescope", *Proc. SPIE* 11447 (2020).
[14] D. Ricci, et al., "Architecture of the SOXS instrument control software", *Proc. SPIE* 10707, 107071G (2018).
[15] OPC foundation website: https://opcfoundation.org/
[16] D. Ricci, et al., "Development status of the SOXS instrument control software", *Proc. SPIE* 11452 (2020).
[17] M. Aliverti, et al., "The mechanical design of SOXS for the NTT", *Proc. SPIE* 10702, 1070231 (2018).
[18] M. Aliverti, et al., "Manufacturing, integration and mechanical verification of SOXS", *Proc. SPIE* 11447 (2020).